\documentclass[usenatbib]{mn2e}
\usepackage{natbibmnfix,graphicx,times, amsmath}

\newcommand{\avenf}{\bar{x}_{\rm HI}}

\newcommand{\lya}{Ly$\alpha$}
\newcommand{\lyb}{Ly$\beta$}

\newcommand{\Msun}{M_\odot}
\newcommand{\Tvir}{T_{\rm vir}}
\newcommand{\fcoll}{f_{\rm coll}({\bf x}, z, R)}

\newcommand{\Mmin}{M_{\rm min}}

\newcommand{\apjl}{ApJ}
\newcommand{\apjs}{ApJS}
\newcommand{\aap}{A\&A}

\newcommand{\taueff}{\tau^{\rm eff}_{\rm GP}}

\newcommand{\taumax}{\tau_{\rm max}}
\newcommand{\dPtot}{P^{\rm dark}_{\rm tot}(<R, \avenf, z, \taumax)}
\newcommand{\dPpost}{P^{\rm dark}_{\rm HII}(\taumax, z)}
\newcommand{\dPpre}{P^{\rm dark}_{\rm HI}(<R, \avenf, z)}
\newcommand{\dPexc}{\Delta P^{\rm dark}(<R, \avenf, z, \taumax)}
\newcommand{\lyg}{Ly$\gamma$}

\def\myputfigure#1#2#3#4#5%
{\vskip#5pt\makebox[0pt]{\hskip#2in
\includegraphics[width=#3\textwidth]{#1}}\vskip#4pt\hfill}

\newcommand\lsim{\mathrel{\rlap{\lower4pt\hbox{\hskip1pt$\sim$}}
        \raise1pt\hbox{$<$}}}
\newcommand\gsim{\mathrel{\rlap{\lower4pt\hbox{\hskip1pt$\sim$}}
        \raise1pt\hbox{$>$}}}

\begin{document}

%\submitted{Submitted to the MNRAS}

\title[Was reionization complete by $z\sim$ 5--6?]{Was reionization complete by $z\sim$ 5--6?}

\author[Mesinger]{Andrei Mesinger\thanks{Hubble Fellow; email: mesinger@astro.princeton.edu} \\
Department of Astrophysical Sciences, Princeton University, Princeton, NJ 08544, USA}
\voffset-.6in

\maketitle

\begin{abstract}
  It is generally taken for granted that reionization has completed by $z=6$, due to the detection of flux in the \lya\ forest at redshifts $z<6$.  However, since reionization is expected to be highly inhomogeneous, much of the spectra pass through just the ionized component of the intergalactic medium (IGM) even for non-negligible values of the volume-weighted mean neutral hydrogen fraction, $\avenf$.  We study the expected signature of an incomplete reionization at z$\sim$ 5--6, using very large-scale (2 Gpc) seminumeric simulations.  We find that ruling out an incomplete reionization is difficult at these redshifts since: (1) quasars reside in biased regions of the ionization field, with fewer surrounding HI patches than implied by the global mean, $\avenf$; this bias extends tens of comoving megaparsecs for $\avenf\lsim0.1$; (2) absorption from the residual neutral hydrogen inside the ionized IGM generally dominates over the absorption from the remaining HI regions, allowing them to effectively ``hide'' among the many dark spectral patches; (3) modeling the \lya\ forest and its redshift evolution even in just the ionized IGM is very difficult, and nearly impossible to do a priori.  We propose using the fraction of pixels which are dark as a simple, nearly model-independent upper limit on $\avenf$.  Alternately, the size distribution of regions with no detectable flux (dark gaps) can be used to place a more model dependent constraint.  Either way, the current sample of quasars is statistically insufficient to constrain $\avenf$ at $z\sim6$ to even the 10 per cent level.  At $z\sim5$, where there are more available sightlines and the forest is less dark, constraining $\avenf\lsim0.1$ might be possible given a large dynamic range from very deep spectra and/or the \lyb\ forest.
We conclude with the caution against over-interpreting the observations.  There is currently no direct evidence that reionization was complete by $z\sim$ 5 -- 6.
\end{abstract}


\section{Introduction}
\label{sec:intro}

Theoretical arguments suggest that the first astrophysical objects ignited at redshifts $z \lsim$ 30--40.  The ionizing radiation of these early stars and black holes carved out bubbles of ionized hydrogen around them.  As subsequent generations of objects formed, these HII regions gradually merged to create a complex HII morphology.  This epoch of reionization terminated in so-called ``overlap'', when the myriad ionization fronts merged, and the only remaining pockets of hydrogen with a substantial neutral fraction were dense, self-shielded systems, such as Lyman limit systems (LLSs) and damped Lyman alpha systems (DLAs).

Reionization is exceedingly complex, and involves uncertain, observationally-underconstrained astrophysics.  Early interpretations of observations generally made the simplifying assumption that reionization is spatially homogeneous.  However, an extremely inhomogenous reionization seems to be an unavoidable conclusion from recent analytic models \citep{FZH04}, as well as large-scale numeric (e.g. \citealt{CSW03, Sokasian03, Iliev06_sim, McQuinn07, TC07}), and seminumeric \citep{Zahn07, MF07, GW08, Alvarez08, Thomas09} simulations.  The patchiness of reionization has many important consequences on the interpretation of observations, prompting calls for caution when interpreting \lya\ damping wing absorption in high-$z$ spectra \citep{MF08damp, McQuinn08}, redshift evolution of the Lyman alpha emitter (LAE) luminosity function \citep{FZH06, McQuinn07LAE, MF08LAE, Iliev08}, the apparent size of the proximity region in high-$z$ quasar spectra \citep{Lidz07, AA07}, and a sharp redshift evolution in the effective Gunn-Peterson (GP; \citealt{GP65}) optical depth, $\taueff$, \citep{FM09}.

Despite its complexity, a hard lower limit for the overlap phase of reionization of $z\sim6$ is often quoted in the literature.  This result is justified by the detection of flux in the \lya\ GP throughs of $z\lsim6$ quasars (QSOs).  Because of the high resonance optical depth of the \lya\ line, the detection of flux implies that the HI fraction in the intergalactic medium (IGM) is less than $\sim10^{-4}$ at these redshifts  (e.g. \citealt{Fan01, Becker01, Djorgovski01, SC02}).  Although this number might be appropriate for the residual HI in the {\it ionized component} of the IGM, when viewed in the context of a patchy reionization the picture becomes less clear; in fact, the step from ``some flux is detected'' to ``overlap has occurred'' might require a large leap of (reionization model based) faith.

The interpretation of the QSO \lya\ forest in this regime is complicated for several reasons.  Firstly, as reionization nears completion, the remaining neutral patches become increasingly rare.  These neutral patches also have very large line-of-sight (LOS) fluctuations, which are larger in 1D than in the analogous 3D volume (c.f. \citealt{LOF06} for a related treatment of the sightline-to-sightline fluctuations arising from just the density field).  Large sections of the line-of-sight therefore should pass through ionized IGM even for non-negligible values of the volume averaged hydrogen neutral fraction, $\avenf\lsim0.1$. 

 Further complicating matters is the fact that QSO host halos are extremely biased, and preferentially reside close to the centers of large HII regions, or large scale regions with fewer HI patches than implied by the global mean neutral fraction (e.g. \citealt{Lidz07, AA07, GW08}).  Thus, one would have to look at spectral regions distant from the emission line center to get a representative sample of the ionization state of the Universe.

 At these high redshifts ($z\sim$ 5--6), the GP troughs are already very dark with large sightline-to-sightline fluctuations {\it even from the residual neutral hydrogen in HII regions assuming a uniform UV background} (UVB; e.g. \citealt{MHR00, Fan06, LOF06}).  Therefore, it is quite challenging to statistically distinguish in the \lya\ spectra the {\it additional} dark patches arising from pre-overlap neutral regions, from the dark patches already present due to the residual neutral hydrogen, LLSs and DLAs in the ionized IGM.  Indeed, \citet{Lidz07} point out that the final overlap stage of reionization might have occurred at $z<6$, given these difficulties in interpreting the observed spectra.  Here we quantify and further explore this scenario.

Specifically, {\it the purpose of this paper is to quantify how confident can we be from current \lya\ forest data that overlap did not occur at $z<6$, given that reionization is very inhomogenous}.  For this purpose, we make use of the seminumeric simulation, DexM\footnote{http://www.astro.princeton.edu/$\sim$mesinger/DexM/}, which allows us to efficiently generate density, halo, and ionization fields on very large scales and with a large dynamic range.  This is an essential property of reionization studies, especially those involving rare, massive QSOs, since one must be able to statistically capture the ionization field, resolve $M\gsim$ $10^{12}$--$10^{13}$ $\Msun$ QSO host halos, while including ionizing photons from very small halos capable of efficient atomic hydrogen cooling ($\Tvir \gsim 10^4$ K, or $M \gsim 10^8 \Msun$ for the redshifts in question).

In \S \ref{sec:sims}, we briefly summarize the large-scale seminumeric simulations we use for this study.  In \S \ref{sec:bias}, we show how QSO host halos locations are biased with respect to the ionization field.  In \S \ref{sec:excess}, we present simple, general estimates of the excess absorption in the \lya\ forest from incomplete, patchy reionization.  In \S \ref{sec:dark_gaps}, we include the geometry of HI patches, and present distributions of spectral dark gaps.  Finally in \S \ref{sec:conc}, we summarize our conclusions.

We quote all quantities in comoving units.  We adopt the background cosmological parameters ($\Omega_\Lambda$, $\Omega_{\rm M}$, $\Omega_b$, $n$, $\sigma_8$, $H_0$) = (0.72, 0.28, 0.046, 0.96, 0.82, 70 km s$^{-1}$ Mpc$^{-1}$), matching the five--year results of the {\it WMAP} satellite \citep{Dunkley09, Komatsu09}.

Throughout the paper, we refer to overlap as the ``end'' of reionization.  As mentioned earlier, the end of reionization is somewhat difficult to define, since even after overlap a few percent of the {\it mass weighted} HI fraction remains in dense absorption systems such as LLSs, DLAs and galaxies (e.g. \citealt{PW09}).  These systems generally correspond to overdensities (e.g. \citealt{MHR00}), unlike the HI patches pre-overlap, which preferentially reside in voids due to the ``inside-out'' nature of reionization on large scales.  Furthermore, these absorption systems are fairly small (e.g. \citealt{Schaye01}) occupying a modest fraction of the total volume of the Universe.
Here we use the {\it volume weighted} neutral hydrogen fraction, $\avenf$, as a proxy for the reionization status.  Thus we define ``post-overlap'' as the regime where $\avenf \rightarrow 0$, i.e. when the only remaining neutral regions are in dense, self-shielded systems.

\section{Seminumerical Simulations}
\label{sec:sims}

\begin{figure*}
\vspace{+0\baselineskip}
{
\includegraphics[width=0.5\textwidth]{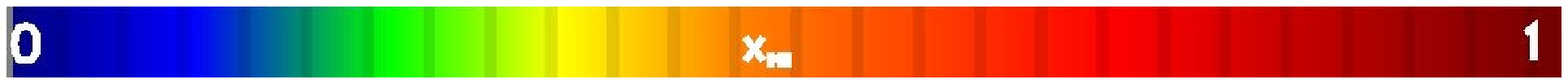}\\
\includegraphics[width=0.33\textwidth]{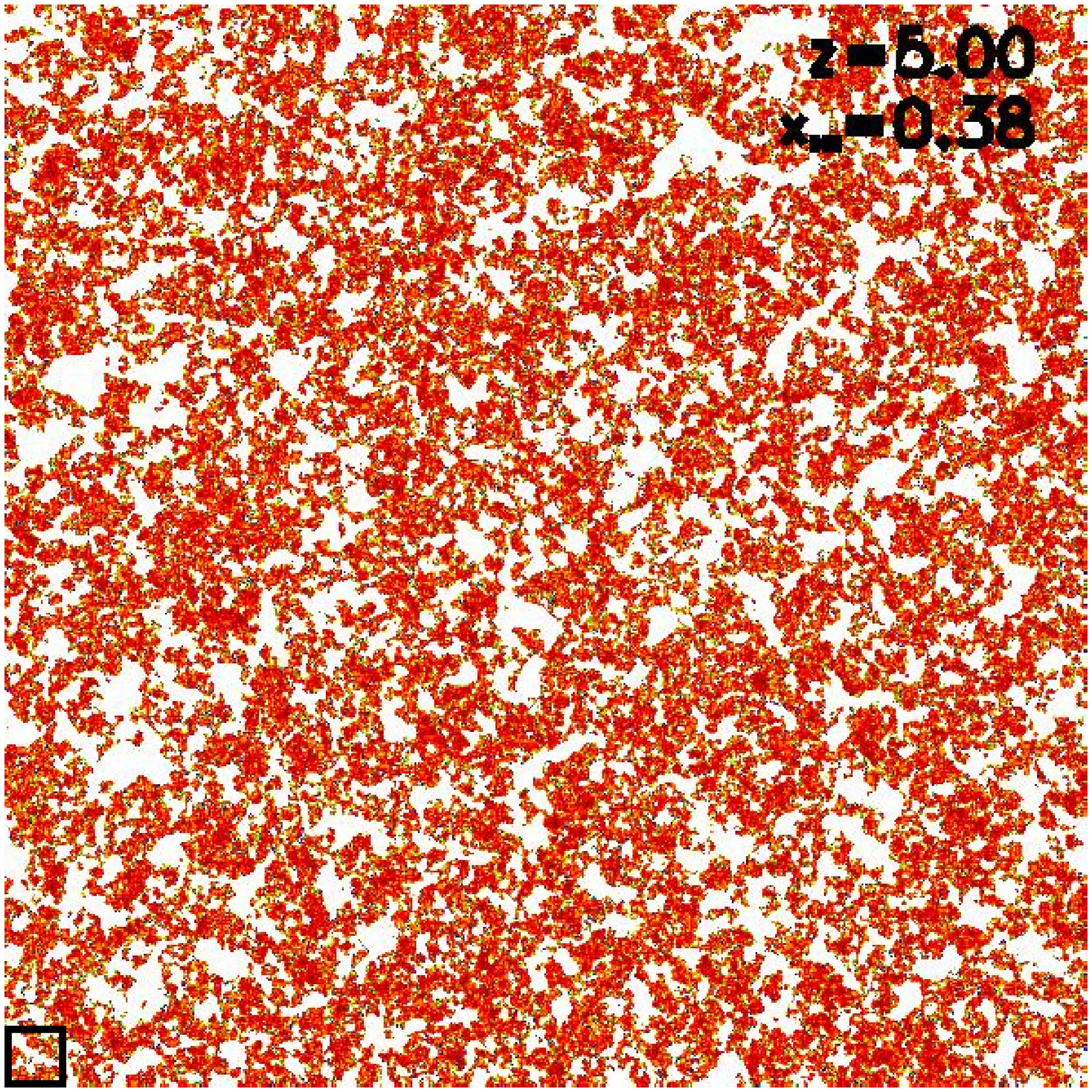}
\includegraphics[width=0.33\textwidth]{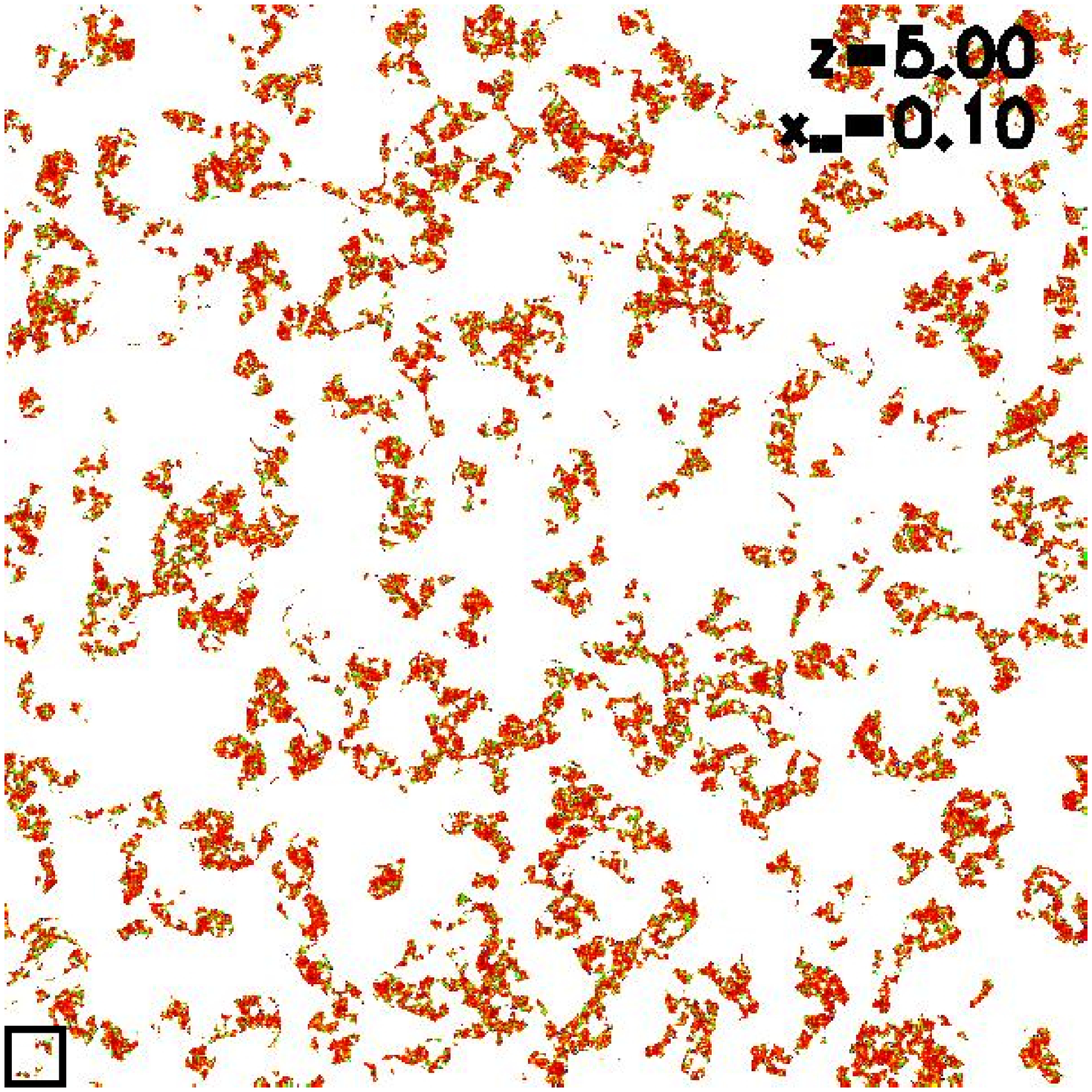}
\includegraphics[width=0.33\textwidth]{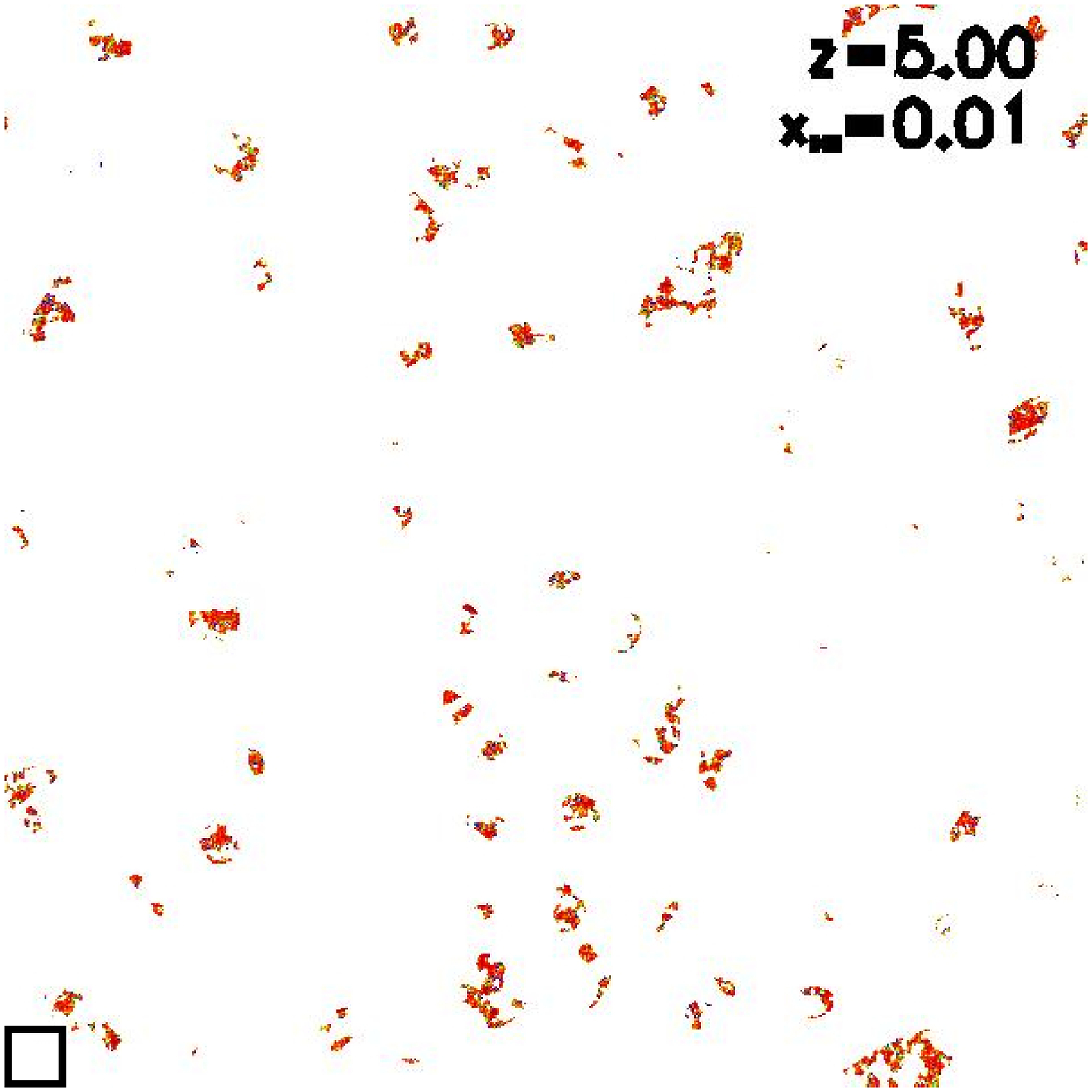}
}
\caption{
Ionization fields at $z=5$, and $\avenf=$ 0.38, 0.10, 0.014, left to right.  Ionized regions are shown in white.  Each slice is 2 Gpc on a side and 3.3 Mpc deep.  The square in the lower left of each panel frames a 100 Mpc $\times$ 100 Mpc region, roughly representative of the maximum size achievable by present-day numerical simulations which resolve halos down to the atomic cooling threshold.
\label{fig:ionization_field}}
\vspace{-1\baselineskip}
\end{figure*}

We use the seminumerical code DexM to generate evolved density, halo and ionization fields at $z=$ 5 and 6.  The halo and density schemes are presented in \citet{MF07}, to which we refer the reader for details.  We use a new algorithm presented in Zahn et al. (in preparation) to generate ionization fields on extremely large scales without having to resolve the ionizing sources (see \citealt{Zahn07, MF07, GW08, Alvarez08, Thomas09} for alternate seminumeric algorithms to generate 3D ionization fields).  Here we briefly outline these simulations.

Our simulation box is 2 Gpc on a side, with the final ionization fields having grid cell sizes of 3.3 Mpc.  Halos are filtered out of the 1800$^3$ linear density field using excursion-set theory.  Halo locations and the initial linear density field are then mapped to Eularian coordinates at a given redshift using first-order perturbation theory \citep{ZelDovich70}.  This approach is similar in spirit to the ``peak-patch'' formalism of \citet{BM96_algo}.  The resulting halo fields have been shown to match both the mass function and statistical clustering properties of halos in N-body simulations, well past the linear regime (\citealt{MF07}, Mesinger et al., in preparation).

We use a new option in DexM which allows us to generate ionization fields directly from the evolved density field; thus we are able to account for the ionizing contribution of halos which are too small to be resolved by our halo finder.  We use excursion-set formalism \citep{FZH04} to flag fully ionized cells in our box as those which meet the criteria $\fcoll \geq \zeta^{-1}$, where $\zeta$ is some efficiency parameter and $\fcoll$ is the collapse fraction smoothed on decreasing scales, starting from a maximum $R_{\rm max}=50$ Mpc and going down to the cell size, $R_{\rm cell}$.  This maximum scale is chosen to roughly match the extrapolated ionizing photon mean free path at these redshifts (e.g. \citealt{Storrie-Lombardi94, Miralda-Escude03, CFG08}).  However, we use the conditional Press-Schechter (PS) formalism (\citealt{LC93, SK99}; specifically we use the ``hybrid'' form of \citealt{BL04}) on the {\it evolved} (i.e. non-linear) density field to calculate $\fcoll$.  Additionally, we allow for partially-ionized cells by setting the cell's ionized fraction to $\zeta f_{\rm coll}({\bf x}, z, R_{\rm cell})$ at the last filter step for those cells which are not fully ionized.  We also account for Poisson fluctuations in the halo number, when the mean collapse fraction becomes small, $f_{\rm coll}({\bf x}, z, R_{\rm cell}) \times M_{\rm cell} < 50\Mmin$, where $M_{\rm cell}$ is the total mass within the cell and our faintest ionizing sources correspond to a halo mass of $\Mmin=2\times10^8 \Msun$.  This last step is found to be somewhat important when the cell size increases to $\sim 1$ Mpc. We generate ionization fields corresponding to several values of $\avenf$ at each redshift by varying our efficiency parameter. Our ionization field algorithm is described in detail in Zahn et al. (in preparation) and found to be in good agreement with cosmological radiative transfer algorithms, though we caution that tests were not yet preformed at the $\avenf\lsim0.01$ level.

We show slices through our ionization fields in Fig. \ref{fig:ionization_field}, for $\avenf=$ 0.38, 0.10 and 0.014, left to right.  Note that the neutral IGM is distributed quite inhomogeneously even at $\avenf=0.38$, with the maximum region of influence, $R_{\rm max}=50$ Mpc, damping the inhomogeneities on large scales.  The 100 Mpc $\times$ 100 Mpc square in the lower right corner of the slices is roughly representative of the maximum size achievable with numerical simulations, if they wish to resolve $\sim10^8 \Msun$ halos which are expected to provide the bulk of ionizing photons.  The square illustrates the power of the seminumerical approach.

After creating the halo, density and ionization fields in our 2 Gpc box, we extract randomly-oriented LOSs of length 500 Mpc originating from $M\gsim 3\times 10^{12} \Msun$ host halos.  This mass range should roughly correspond to the observed high-$z$ QSO host halo masses \citep{WL03, Croton09}, given that their black hole masses are estimated to be $M_{\rm bh} \sim$ $10^8$--$10^9$ $\Msun$ \citep{Kurk07}.  We conservatively\footnote{Here and throughout the paper, we shall use ``conservative'' to mean closer to the average global ionization state and with less scatter than expected.  Thus ``conservative'' choices make the detection of pre-overlap at $z<6$ easier, in our application.  In this case, by generating the same number of LOSs for each halo, the typical LOS originates from the less massive, less biased halos on the low end of the mass range.} extract 500 LOSs per halo, yielding a total of $\sim$ $8\times10^5$ ($10^7$) sightlines at $z=$ 6 (5).

Below we discuss a few simple statistics concerning these LOSs.  Specifically, we are interested in whether the current 22 (101) published QSO spectra at $z>$ 6 (5) are sufficient to distinguish a pre-overlap from a post-overlap IGM.  We stress that we do not create detailed mock spectra for these purposes and take into account only absorption at the \lya\ line center.  Instead, we approach the issue from a more general standpoint and try to draw simple conclusions with minimal model dependence.
Since we are not making precise comparisons to observations, detailed mock spectra are both unnecessary and impossible to create given the comparatively low resolution of our boxes. Furthermore, neglected effects such as the damping wing absorption should not have a large impact on our conclusions, since the absorption profile in the $\avenf\ll1$ regime is on average weak and steep (c.f. Fig. 8 in \citealt{MF08damp}), and so should not have a large impact on many neighboring cells, given their large (3.3 Mpc) size.

\section{Results}
\label{sec:results}

\subsection{Bias of QSOs with respect to the ionization field}
\label{sec:bias}

\begin{figure}
\vspace{+0\baselineskip}
\myputfigure{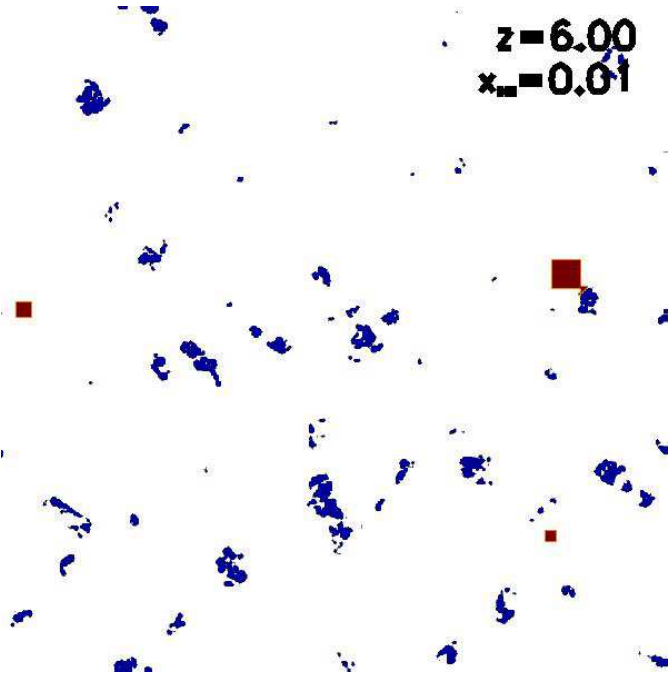}{3.3}{0.5}{.}{0.}
\caption{
Slice through our simulation box at $z=6$, $\avenf=0.01$.  Partially neutral cells are shown in blue; quasar host halos are shown as red squares (enlarged for viewing purposes with the length of a side proportional to the log of the halo mass).  The slice is 6.67 Mpc thick and 2 Gpc on a side.
\label{fig:slice}}
\vspace{-1\baselineskip}
\end{figure}

In Fig. \ref{fig:slice} we show a 2 Gpc $\times$ 2 Gpc $\times$ 6.7 Mpc slice through our simulation box, at $(z, \avenf)=$ (6, 0.01).  The neutral regions are marked in blue, and the quasar host halos are shown as red squares, with the length of the square proportional to the log of the halo mass.  From the picture, one can qualitatively see that larger clusters of HI regions are preferentially farther away from the host halos.  In our pre-overlap model, such HI patches are generally found in large-scale underdense regions or voids, since reionization proceeds ``inside-out'' on large scales.

 We note that we conservatively do not include the ionizing radiation from the quasars in constructing the ionization field.  The ionization field is computed directly from the evolved density field, using the expected (mean) collapse fraction + Poisson scatter due to discrete sources.  Since this collapse fraction corresponds to an integral over a steep halo mass function, it is dominated by the abundance of sources much less massive than QSO host halos.  One could in principle post-process the ionization fields and ``paint-on'' the QSO contribution around the QSO host halos, but this would involve many uncertain assumptions about QSO properties.  Not including the QSO ionizing radiation in the ionization fields is a conservative assumption, because it decreases how ``over-ionized'' or biased their environments are.

\begin{figure}
\vspace{+0\baselineskip}
\myputfigure{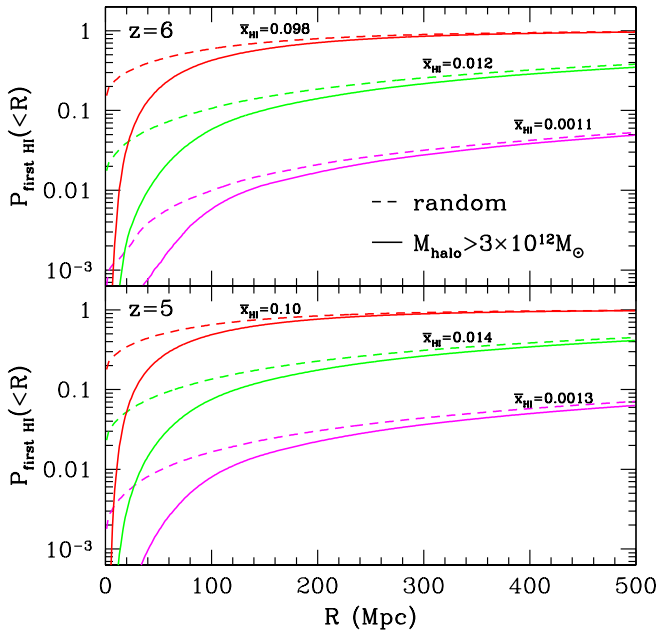}{3.3}{0.5}{.}{0.}
\caption{
Cumulative probability distributions that a LOS will encounter a partially neutral region at distances less than R, originating from $M>3\times10^{12}\Msun$ halos ({\it solid curves}) and random locations ({\it dashed curves}).  The top panel corresponds to $z=6$ and the bottom panel corresponds to $z=5$.  Pairs of curves are drawn from simulation boxes with $\avenf=$ 0.1, 0.01, and 0.001, top to bottom.
\label{fig:pRs}}
\vspace{-1\baselineskip}
\end{figure}

We briefly quantify this bias in Fig. \ref{fig:pRs}, which shows the cumulative probability distributions that a LOS will encounter a partially neutral region at distances less than R.  The top panel corresponds to $z=6$ and the bottom panel corresponds to $z=5$.  Pairs of curves are drawn from simulation boxes with $\avenf=$ 0.1, 0.01, and 0.001, top to bottom.  The solid curves originate from our QSO host halos, while the dashed curves originate from random locations.

There are two important things to note from this figure.  Firstly, we confirm quantitatively that reionization, especially as probed by 1D skewers, is extremely patchy.  Most of the LOSs for $\avenf\lsim0.01$ {\it do not intersect a single neutral region in 500 Mpc}.  To put this into perspective, 500 Mpc corresponds to the entire redshift interval from $z=$ 6 to 5.  Thus in this regime, a large majority of the \lya\ forest sightline pre-overlap goes through ionized IGM, and is by definition indistinguishable from sightlines going through a post-overlap IGM.

Secondly, we see that the likelihood of encountering the first neutral patch within the first $\sim$100 Mpc is significantly smaller for those sightlines originating at QSO host halos, than those originating at random locations in the box.  In fact, sightlines originating from QSO halos are over twice as likely to entirely go through HII regions within the first 60--80 Mpc, for $\avenf\lsim0.1$.  Again, we are being conservative in not explicitly including the ionizing radiation from the QSOs, which would make nearby neutral patches even more unlikely. Thus, Fig. \ref{fig:pRs} paints a pessimistic picture about being able to claim to measure $\avenf\lsim0.1$ at this epoch.  We will show the radial dependence of the spectral number density of these HI patches in the following section.

\subsection{Excess Absorption in the Pre-Overlap IGM}
\label{sec:excess}

Perhaps the simplest, least model-dependent upper limit on $\avenf$ can be obtained through the covering factor of dark pixels at a given redshift.  As the \lya\ spectra are roughly half dark at $z\sim5$ (see below and figures 4 in \citealt{Fan02} and 11 in \citealt{BRS07}), this upper limit is not particularly awe-inspiring.  However, it is likely that given a large dynamic range, such as might be available from very deep spectra and the \lyb\ forest, one might be able to place an upper limit of $\avenf\lsim0.1$ (provided that the Universe is indeed so highly ionized).  In this section we explore the possibility of doing better that this simple upper limit.  This entails having some a priori knowledge of the \lya\ absorption in the ionized component of the IGM.  Although very challenging, one could perhaps use the integrated observed UV luminosity function in lieu of extracting the UV background directly from the forest with the a priori assumption that it is in the post-overlap regime.  Such an estimate of the ionizing background inside the ionized IGM component, coupled with a prescription for estimating the residual neutral hydrogen distribution [e.g. radiative equilibrium and/or a simple density threshold for self-shielding \citep{MHR00}], can yield an independent estimate of the absorption inside the ionized IGM, which one can then apply to observed spectra in an effort to detect the excess absorption expected in a pre-overlap IGM.  Unfortunately, astrophysical uncertainties are large at the present time, and this procedure is unlikely to yield strong constraints.

As explained in the introduction, detecting pre-overlap is more difficult than merely having enough LOSs to find one of the neutral patches seen in figures \ref{fig:slice} and \ref{fig:pRs}.  One must be able to statistically distinguish such a dark region in the spectra from those already resulting from the residual HI in the ionized component of the IGM.  As seen in the previous figures, a large majority of the LOS will go through ionized IGM, for these global neutral fractions, making them identical to the post-overlap spectra.  We will now proceed to do some simple estimates on the amount of excess absorption in \lya\ spectra due to the neutral regions in the pre-overlap regime.

  Because we don't know redshift evolution of $\avenf$, the UVB, absorption systems, temperature, etc., the safest route is to isolate the spectral imprint of the pre-overlap HI regions by analyzing spectra at a fixed redshift, instead of binning data in wide redshift bins. We first focus on the number of ``dark'' spectral regions in the spectra pre and post overlap.  We note that the minimum sizes of dark spectral regions are resolution dependent.  The resolution of our simulation is 3.3 Mpc, which is approximately three times the resolution of the Keck ESI spectra at these redshifts, when binned to R=2500 in order to increase the dynamical range (e.g. \citealt{White03, Fan06}).  Thus, our resolution is somewhat larger than the minimum size of a statistically significant detection of a ``dark region'' ($\tau\gsim4$) with present-day instruments.  Note that very deep, high resolution spectra were taken for a few ($\sim9$) of these $z>5$ objects with the Keck HIRES instrument \citep{BRS07}.  These can also be binned to lower resolution in order to increase the dynamical range.  Alternately, one can sacrifice some of the additional dynamical range obtained through binning in order to resolve narrow transmission spikes, which might yield more discriminating dark gap distributions than those discussed in \S \ref{sec:dark_gaps}.

\begin{figure}
\vspace{+0\baselineskip}
\myputfigure{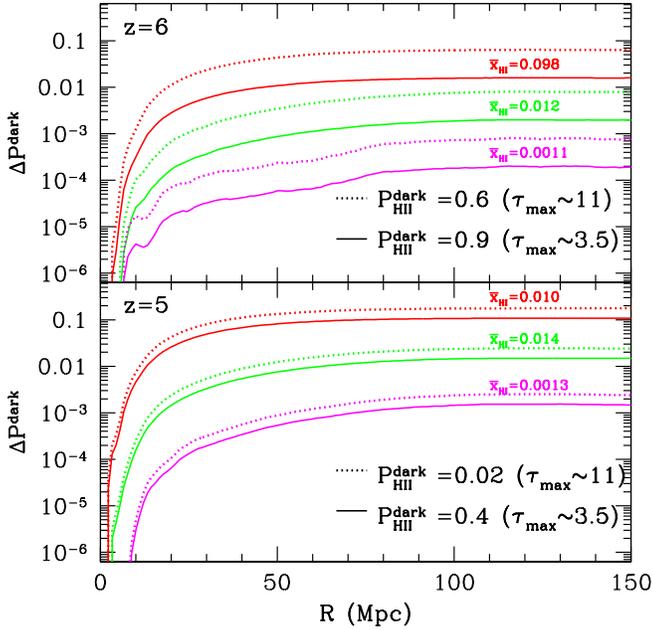}{3.3}{0.5}{.}{0.}
\caption{
The {\it excess} probability of finding a dark spectral region pre-overlap (compared to post-overlap) at distance $R$ away from a quasar, at $z=6$ (top panel) and $z=5$ (bottom panel).  Dotted curves correspond to a post-overlap probability of $\dPpost=0.6$ ({\it top panel}) and $\dPpost=0.02$ ({\it bottom panel}) of encountering a dark region, roughly corresponding to the mean value in the \lyb\ forest.  Solid curves correspond to $\dPpost=0.9$ ({\it top panel}) and 
$\dPpost=0.4$ ({\it bottom panel}), which is roughly equivalent to a maximum Ly$\alpha$ optical depth of $\taumax = 3.5$ ({\it see text}).  Pairs of curves are drawn from simulation boxes with $\avenf=$ 0.1, 0.01, and 0.001, top to bottom.
\label{fig:DeltaPpre}}
\vspace{-1\baselineskip}
\end{figure}

We define $\dPpre$ to be the probability of encountering a neutral region (i.e. {\it not within} the ionized component of the IGM) at a distance between $R$ and $R+dR$ away from a QSO.  Similarly, we define $\dPpost$ to be the post-overlap (i.e. {\it within} the ionized component of the IGM) probability of encountering a dark spectral region with optical depth greater than $\taumax$.  For simplicity of presentation, we ignore the radial dependence of $\dPpost$, which should in general be much weaker than the radial dependence of the ionization field and $\dPpre$.  For our regime of interest ($\avenf<<1$) and neglecting spatial correlations, we can approximate the total probability of encountering a dark pixel at a distance $R$ away from a QSO as:

\begin{equation}
\label{eq:dPtot}
\dPtot = P^{\rm dark}_{\rm HI} + P^{\rm dark}_{\rm HII} - P^{\rm dark}_{\rm HI} P^{\rm dark}_{\rm HII} ~ .
\end{equation}

 Note that these probabilities are different than the ones plotted in Fig. \ref{fig:pRs}, where we had plotted the likelihood of encountering the {\it first} neutral region at distances less than $R$.  As defined, $P^{\rm dark}_{\rm HI}$ and $P^{\rm dark}_{\rm HII}$ are analogous to the covering fraction of dark regions, i.e. the fraction of pixels at fixed $z$ which are dark.

Therefore, the {\it additional} likelihood of finding a dark spectral region in a pre-overlap universe compared to a post-overlap universe is:

\begin{equation}
\label{eq:dPexc}
\dPexc = P^{\rm dark}_{\rm HI} ( 1 - P^{\rm dark}_{\rm HII}) ~ .
\end{equation}

\noindent As $R\rightarrow\infty$ and $\dPpost\rightarrow0$, the likelihood of encountering a dark region roughly approaches the global neutral fraction, $\avenf$.

In Fig. \ref{fig:DeltaPpre}, we plot this excess probability at $z=6$ (top panel) and $z=5$ (bottom panel).  Solid curves correspond to $\dPpost=0.9$ ({\it top panel}) and $\dPpost=0.4$ ({\it bottom panel}).  These were computed from the 150 Mpc simulations in \citet{MF09}, using a minimum source halo mass of $\Mmin=1.6\times10^8 \Msun$ and ionizing photon mean free path of 40 Mpc,  These values serve as a rough estimate of the fraction of the \lya\ spectra with absorption greater than the threshold $\taumax=3.5$, when the UVB is adjusted to match the observed value of $\taueff\equiv -\ln \langle \exp[-\tau] \rangle_{\rm GP}=$ 7.1 (2.1) at $z\sim$ 6 (5) \citep{Fan06}. The dotted curves correspond to a post-overlap probability of $\dPpost=0.6$ ({\it top panel}) and $\dPpost=0.02$ ({\it bottom panel}), corresponding to an increased dynamical range, $\taumax=11$.  This value is (very) roughly the mean detection limit for the \lyb\ forest for the Keck ESI.  Again, the quoted values of $\dPpost$ serve as a guide and should not be taken too seriously as we do not create detailed mock spectra, though they do seem to be in decent agreement with observations (c.f. Fig. 4 in \citealt{Fan02} and Fig. 11 in \citealt{BRS07}).

Fig. \ref{fig:DeltaPpre} further illustrates the biased ionization structure surrounding the QSO host halos.  Even without including QSO ionizing radiation, photons must travel tens of megaparsecs to get to a representative sample of the IGM.  This distance becomes larger for smaller values of $\avenf$.  The scale of this bias can even exceed the local QSO region of influence.  In other words, the galaxies surrounding the quasar can self-ionize (or pre-ionize if they turn on before the QSO as is likely) the surrounding IGM to scales larger than the QSO's proximity zone, for these values of $\avenf$ (c.f. \citealt{AA07}).  We plan to explore this effect further in a subsequent paper, in conjunction with hydrodynamical simulations.  Note however that using this number density statistic, the bias is not as extreme as with the ``first HI patch'' statistic presented in Fig. \ref{fig:pRs}.

The asymptotic value of $\dPexc \sim \avenf (1-P^{\rm dark}_{\rm HII})$ is approached at $R\sim$ 30--60 Mpc.  Strictly speaking, $\dPexc$ can exceed this value somewhat, since most HI regions pre-overlap on this scale are partially ionized to few$\times$10\%.  As the dynamic range probed by the Lyman lines is so narrow, partially ionized regions pre-overlap are counted the same as fully neutral regions in these statistics.

\begin{figure}
\vspace{+0\baselineskip}
\myputfigure{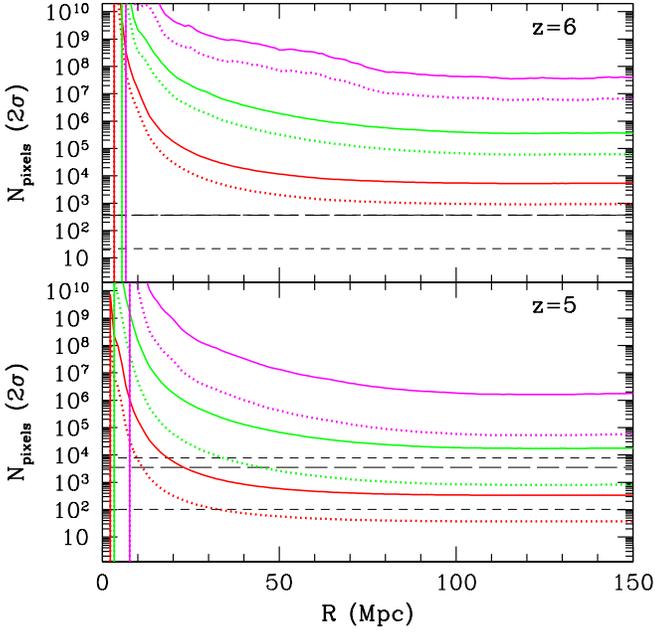}{3.3}{0.5}{.}{0.}
\caption{
The number of $\sim3$ Mpc pixels, $N_{\rm pixels}$, required to separate the pre-overlap and post-overlap expectation values of the number of dark regions by 2$\sigma$.  In other words, $N_{\rm pixels}$ is the minimum value such that an observed number of dark pixels cannot be consistent with {\it both} a pre-overlap and post-overlap IGM, to within 2$\sigma$.  Curves correspond to the same models discussed in Fig. \ref{fig:DeltaPpre}. $N_{\rm pixels}$ scales as n$^2$, where n is the number of standard deviations separating the two distributions.  The lower horizontal short-dashed lines demarcate the present number of published QSO spectra at $\geq z$.  The upper horizontal short-dashed lines demarcate the total number of $\sim3$ Mpc pixels/segments available from published quasar spectra at $\geq z$.  The horizontal long-dashed lines show this total for the \lyb\ forest.
\label{fig:Nsamples}}
\vspace{-1\baselineskip}
\end{figure}

Before overlap, the expectation value of the number of dark pixels at a given redshift from a total of $N_{\rm pixels}$ samples is $N_{\rm pixels} P^{\rm dark}_{\rm tot}$.  Again, $P^{\rm dark}_{\rm tot}$ is the likelihood a pixel will be dark {\it either} from an HI region {\it or} from a dark region within the ionized IGM.  After overlap, when the entire IGM is ionized, the expectation value of the number of dark pixels is smaller: $N_{\rm pixels} P^{\rm dark}_{\rm HII}$.  The PDFs of these two distributions should be binomial with these mean values.  Therefore, in order to separate the expectation values pre and post overlap by $n$ standard deviations, $\sigma$, we require:

\begin{equation}
\label{eq:Npixels}
N_{\rm pixels} P^{\rm dark}_{\rm tot} - n \sigma_{\rm tot} \geq N_{\rm pixels} P^{\rm dark}_{\rm HII} + n \sigma_{\rm HII}
\end{equation}

\noindent  In other words, $N_{\rm pixels}$ is the minimum value such that an observed number of dark pixels cannot be consistent with {\it both} a pre-overlap and post-overlap IGM, to within $n\sigma$.

In Fig. \ref{fig:Nsamples} we plot the quantity $N_{\rm pixels}$ for $n=2$, for the same models shown in Fig. \ref{fig:DeltaPpre}.  The lower horizontal short-dashed lines demarcate the present number of published QSO spectra at $\geq z$.  This limit corresponds to the least model-dependent constraint possible on $\avenf$, as it doesn't involve binning data in wide redshift bins or modeling the redshift evolution of these parameters.

If one is allowed to make assumptions about the redshift evolution of $\avenf$, and the absorption in the ionized IGM, then one can use wider spectral regions, approaching the \lya\ line and the intrinsic source redshift. This is the {\it most} information we can hope to squeeze out from the observed sample of quasar spectra, using this simple statistic.  We show this limit as the upper horizontal short-dashed lines, demarcating the total number of $\sim3$ Mpc pixels/segments in published \lya\ forest spectra available for analysis from quasars at $\geq z$.  The horizontal long-dashed lines show this total for the \lyb\ forest.  Our horizontal lines are also conservative limits in that we count spectral regions all the way to the quasar redshift.  In reality, the biasing seen in the previous figures from the surrounding ionization field, a well as from the QSO's ionizing radiation will make the region within $R\lsim50$ Mpc unusable for this analysis.  Requiring that the quasar is at redshifts greater than $z + 50$ Mpc, eliminates roughly half (quarter) of the total sample at $z=$ 6 (5), though the total LOS length is only weakly affected.

Furthermore, this analysis is extremely conservative in that it assumes perfect a priori knowledge of the absorption caused by the ionized component of the IGM, i.e. $P^{\rm dark}_{\rm HII}$.  Even with such unrealistic confidence in our ability to model the UVB, density, and temperature distributions in the ionized IGM\footnote{This is challenging not only since it is model-dependent, but also since the best estimate of the UVB and temperature comes from the \lya\ forest, with the a priory assumption that it is in the post-overlap regime.  A \lya\ forest in a pre-overlap IGM would bias the derived quantities.},
%  One can decrease this bias with the extended dynamical range provided by the higher Lyman lines.  In other words, one can create better estimates of the absorption in HII regions by making sure that the pixels have flux in the higher-order lyman transitions.  If these are dark as well, that pixel(s) has a higher likelihood of corresponding to an HI region pre-overlap.},
 we see from Fig. \ref{fig:Nsamples} that many LOSs are required to constrain the additional absorption caused by $\avenf\lsim0.1$.  In practice using this covering fraction statistic without a priori knowledge of the absorption inside the ionized IGM, means that a pre-overlap IGM is {\it degenerate} with a ``darker'' post-overlap IGM (e.g. with a weaker UVB).  This means that obtaining constraints on $\avenf$ is much more difficult than implied by Fig. \ref{fig:Nsamples}.

At $z=6$, there is no hope of using the current sample of 22 $z\geq6$ QSOs to constrain $\avenf$ to the $\sim10$\% level, using the mean excess absorption in the \lya\ or \lyb\ forest.  At $z=5$, one could marginally constrain $\avenf\lsim0.1$, from the available data, but only in a model-dependent way.

The \lyb\ forest probes a larger dynamic range, and so can better discriminate between the residual HI in the ionized IGM and the more highly neutral patches pre-overlap (e.g. \citealt{MH04}).  Thus the dotted curves in Fig. \ref{fig:Nsamples} can be significantly lower than the solid curves.  Fig. \ref{fig:Nsamples} suggests that the \lyb\ forest can be used to potentially constrain $\avenf\lsim0.1$ at $z=5$.

The precise interpretation of the \lyb\ forest is hampered by the presence of lower redshift \lya\ forest absorption, contributing fluctuations in the transmitted flux of order $\sim$ 20\% \citep{DLH04}.  Furthermore, the \lyb\ forest (with its red edge corresponding to the onset of the \lyg\ forest) is narrower than the \lya\ forest.  Thus the horizontal long-dashed curves in Fig. \ref{fig:Nsamples} are lower than the short-dashed ones.  Therefore detailed simulated spectra should be employed when comparing with observations to further study the utility of the \lyb\ forest. Nevertheless, we stress that using the covering fraction of dark pixels in the \lyb\ forest likely provides the {\it simplest, least-model dependent upper limit on the value of $\avenf$.}

Finally, we stress that the main conclusions presented in this section are unaffected by the precise geometry of the neutral patches.  In the $R\rightarrow\infty$ asymptotic limit, the covering fraction of the HI patches roughly approaches $\avenf$, and so is not affected by the possible inaccuracies of our ionization field models.

\subsection{Dark Gap Distribution}
\label{sec:dark_gaps}

Thus far we have ignored geometric information imprinted by the pre-overlap absorption features.  The sizes of regions with no detectable flux (``dark gaps'') can potentially provide useful additional information about the state of the IGM (e.g. \citealt{Croft98, SC02, Barkana02}).  The simplest approach in dark gap studies is to associate pre-overlap with larger spectral dark gaps (e.g. \citealt{Fan06} and references therein).  In practice however, obtaining constraints is difficult even under idealized assumptions (e.g. \citealt{GCF06}).  

In order to gain physical insight into the use of this approach and the robustness of its application, one should be able to disentangle redshift, $\avenf$ (which depends on the large-scale ionization history), and the absorption inside HII regions (which in turn depends on the UVB, density and temperature fields).  Numerical simulations generally relate all of these quantities via some star formation prescription, resulting in a single realization of a possible reionization history.  This lack of versatility makes it difficult to robustly isolate the imprint of patchy reionization from mock spectra.  Further complicating matters are the large parameter space to explore and the enormous dynamic range, as simulations should statistically sample the large-scale environments of QSOs (with halo masses of $\sim10^{13} \Msun$), while resolving the small mass halos dominating the ionizing photon budget (with halo masses of $\sim10^8 \Msun$).  Therefore it is quite useful to extend the available dynamic range through seminumerical simulations such as ours, or sub-grid numerical simulations \citep{KGH07}.

Here we take a more general approach, by separately studying the dark gaps from HI patches and those from residual neutral hydrogen in HII regions. We vary vary $z$, $\avenf$, and the mean GP absorption from the ionized IGM, parametrized by its $\taueff$.  Our poor resolution prevents us from probing the dark gap distribution on small scales, but present technology makes these scales difficult to observe anyway, since spectra need to be binned to increase the dynamical range (c.f. \citealt{White03}).

\begin{figure}
\vspace{+0\baselineskip}
\myputfigure{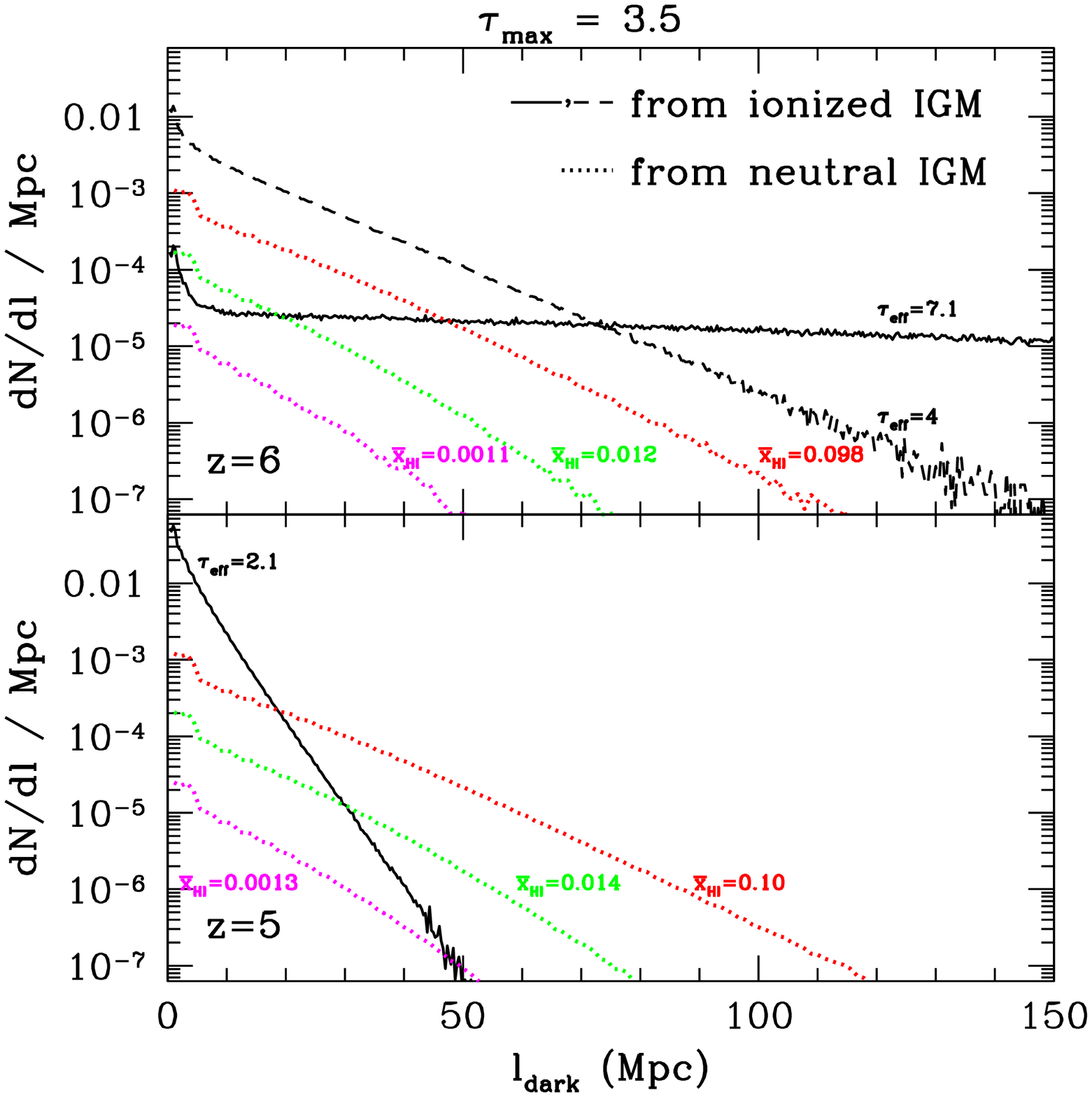}{3.3}{0.5}{.}{0.}
\caption{
The number of dark gaps per unit gap length per spectral Mpc, at $z=6$ ({\it top}) and $z=5$ ({\it bottom}).  Solid and dashed curves are extracted from the post-overlap ionized IGM in the seminumerical simulations of \citet{MF09} assuming $\taumax=3.5$.  The dashed curve corresponds to an effective GP optical depth through entirely ionized IGM of $\taueff=4$, and solid curves correspond to the observed values of $\taueff=7.1$ ({\it top panel}) and $\taueff=2.1$ ({\it bottom panel}).  Dotted curves show the contribution from neutral IGM patches pre-overlap, and correspond to $\avenf=$ 0.1, 0.01, and 0.001, right to left. The total pre-overlap dark gap PDF is approximately the sum of the distributions in the ionized and neutral IGM.
\label{fig:dark_gaps}}
\vspace{-1\baselineskip}
\end{figure}

\begin{figure}
\vspace{+0\baselineskip}
\myputfigure{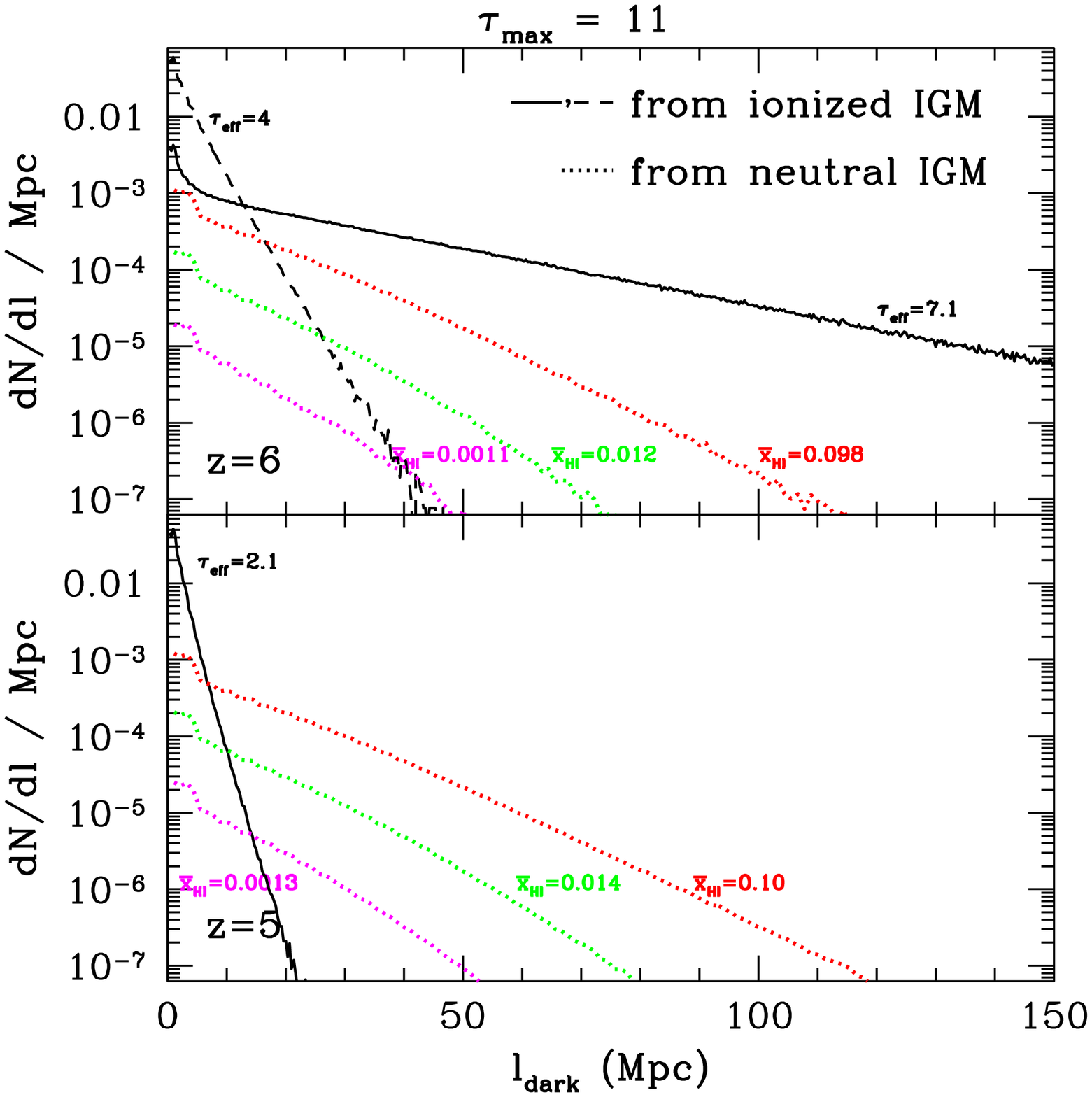}{3.3}{0.5}{.}{0.}
\caption{
Analogous to Fig. \ref{fig:dark_gaps}, but with $\taumax=11$.
\label{fig:dark_gaps_beta}}
\vspace{-1\baselineskip}
\end{figure}

\begin{figure}
\vspace{+0\baselineskip}
\myputfigure{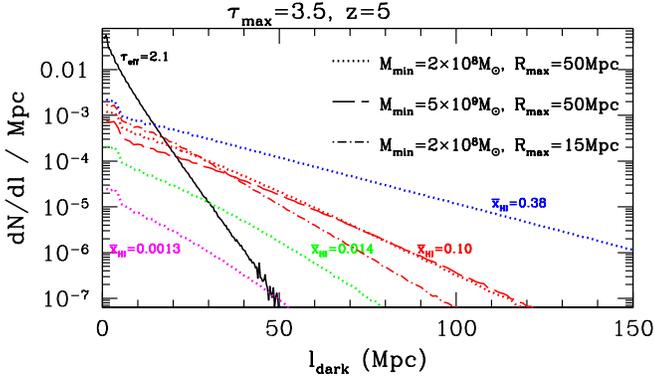}{3.3}{0.5}{.}{0.}
\caption{
Analogous to the bottom panel of Fig. \ref{fig:dark_gaps_beta}, but including some additional models of the ionization field.  The dotted curves show our fiducial model with $\Mmin=2\times10^8 \Msun$ and $R_{\rm max} = 50$ Mpc, with the added blue curve corresponding to $\avenf=0.38$.  There are two additional distributions for $\avenf=0.1$: the dashed curve corresponds to $\Mmin=5\times10^9 \Msun$ and $R_{\rm max} = 50$ Mpc and the dot-dashed curve corresponds to $\Mmin=2\times10^8 \Msun$ and $R_{\rm max} = 15$ Mpc.
\label{fig:additional_dark_gaps}}
\vspace{-1\baselineskip}
\end{figure}

In Fig. \ref{fig:dark_gaps}, we plot the number of dark gaps per unit gap length per comoving Mpc of the LOS, at $z=6$ ({\it top}) and $z=5$ ({\it bottom}).  Solid and dashed curves are extracted from the post-overlap ionized IGM in the seminumerical simulations of \citet{MF09} assuming $\taumax=3.5$.  The dashed curve corresponds to $\taueff=4$, and solid curves correspond to the observed\footnote{Note that here we define $\taueff$ only with respect to the ionized IGM. The ``true'' $\taueff$ in the \lya\ forest pre-overlap can have a contribution from absorption in HI regions.} values of $\taueff=7.1$ ({\it top panel}) and $\taueff=2.1$ ({\it bottom panel}) \citep{Fan06}.  Dotted curves show the contribution from neutral IGM patches pre-overlap, and correspond to $\avenf=$ 0.1, 0.01, and 0.001, right to left. The total pre-overlap dark gap PDF is approximately the sum of the distributions in the ionized and neutral IGM.

Interestingly, Fig. \ref{fig:dark_gaps} shows that the dark gap PDF arising from the neutral patches pre-overlap is insensitive to redshift over this range.  The absolute normalizations scale with $\avenf$, but the shapes of these PDFs are fairly constant.

The dark gap distribution from the ionized IGM has a more complex behaviour.  We see that as the \lya\ forest in the ionized IGM gets darker, the distribution of dark gaps flattens.  The observed (going through ionized and potentially also neutral IGM) GP effective mean optical depth evolves from $\taueff = 2.1$ at $z\sim5$ to $\taueff = 7.1$ at $z\sim6$ \citep{Fan06}.  These values set upper limits to the amount of absorption coming from just the ionized phase of the IGM.  Hence, the true dark gap distribution in the ionized IGM should be similar to or steeper than the solid curves in Fig. \ref{fig:dark_gaps}.

At $z=5$ the dark gap PDF from the ionized IGM is steeper than the PDF from the neutral IGM.  In this regime, a high-value tail in an observed dark gap distribution would indeed be an indication of incomplete reionization.  At $z=6$ however, the situation is more ambiguous.  Here the dark gap distribution from the ionized IGM can be flatter than the distribution from the neutral IGM.  In this regime, a high-value tail in observed spectra could result from {\it either} the ionized component {\it or} the neutral component.  Throughout this redshift range, a signature of pre-overlap seems to be a ``knee'' in the total observed dark gap distribution, as the PDFs from the ionized and neutral components are bound to have different slopes within $5<z<6$.

Therefore at first glance, Fig. \ref{fig:dark_gaps} hints at two signatures of incomplete reionization: (1) the presence of a high-value ($l_{\rm dark} \gsim$ tens of megaparsecs) tail in the observed dark gap distribution at $z\lsim5$; and (2) a redshift-independent signature of a knee in the dark-gap distribution.  Do we have enough data to statistically extract these signatures? 

At $z>6$, there are a total of 22 published QSO spectra, with a total comoving path length of $\sim1200$ Mpc.  Looking at the top panel of Fig. \ref{fig:dark_gaps}, this is likely one to two orders of magnitude too small to create an accurate dark gap PDF.  Furthermore, most QSOs are at redshifts close to $z\sim6$, thus incapable of probing the high-value tail, even if their near zones and ionization field bias could be included in the analysis.  The dark gap PDF from just the ionized IGM is already so flat that large gaps on cosmological scales are likely.  Therefore redshift evolution must be taken into account (see below).

As $z>5$, there are a total of 101 published QSOs, with a total of $\sim 26$ Gpc of combined spectral length of the \lya\ forest. Even excising the closest $\sim50$ Mpc to each quasar yields $\gsim20$ Gpc with which to probe the dark gap PDF.  Comparing to the bottom panel of Fig. \ref{fig:dark_gaps}, this seems like a decently large number, allowing one to constrain $\avenf\lsim0.1$ from the dark gap PDF.

However, in order to take advantage of the available spectral path lengths, one must average over the wide redshift intervals these spectra span.  The resulting total dark gap PDF will absorb the uncertain redshift evolution of both the dotted and solid curves in Fig. \ref{fig:dark_gaps}.  As LOS skewers piercing the ionized IGM get rapidly darker at $z>5$, their corresponding contribution to the total dark gap PDF will get rapidly flatter with redshift, even over fairly small redshift bins. The resulting redshift-averaged PDFs will likely smear out any of the signatures of incomplete reionization discussed above.  Thus to get constraints on $\avenf$ even at the 10\% level, one would again need to rely heavily on our ability to accurately model the absorption from residual HI in the ionized IGM, {\it and} its redshift evolution.

The situation improves with a larger dynamic range, shown in Fig. \ref{fig:dark_gaps_beta} where we plot the same quantities, but taking $\taumax=11$.  This value of $\taumax$ roughly corresponds to an optimistic prediction for the \lyb\ forest, provided that the lower redshift \lya\ forest can be statistically extracted.  We see that the dark gap distributions from an ionized IGM are steeper, thereby increasing the confidence at which one can associate a long dark gap with incomplete reionization at $z\sim5$.  An increased dynamic range might be available with new instruments such as the James Webb Space Telescope ({\it JWST}) and the ten meter telescope (TMT), or through long-integration with existing instruments such as Keck HIRES \citep{BRS07}, 
%AM: (look up specs...other near infrared telescopes),
 or with higher-order Lyman lines, such as \lyb. Although an extended dynamical range is more readily available through the \lyb\ forest in existing spectra, one must be careful in modeling the fluctuating absorption from the superimposed lower redshift \lya\ forest.

In Fig. \ref{fig:additional_dark_gaps}, we plot several additional dark gap distributions at $z=5$, varying two parameters of our ionization models. The dotted curves show our fiducial model with $\Mmin=2\times10^8 \Msun$ and $R_{\rm max} = 50$ Mpc, with the added blue curve corresponding to $\avenf=0.38$.  We see that this curve is noticeably flatter than the curves corresponding to lower neutral fractions.  There are two additional distributions for $\avenf=0.1$: the dashed curve corresponds to $\Mmin=5\times10^9 \Msun$ (as might be the case if feedback is very strong) and $R_{\rm max} = 50$ Mpc while the dot-dashed curve corresponds to $\Mmin=2\times10^8 \Msun$ and $R_{\rm max} = 15$ Mpc [as might be the case if recombinations act as efficient photon sinks in ionized regions (e.g. \citealt{FO05})].  We see that the scenarios in which the sources are more homogeneous (lower $\Mmin$) and more localized (lower $R_{\rm max}$), result in a more uniform (i.e. steeper) distribution of dark gaps.  However, the differences in the curves at fixed $\avenf$ are relatively small, especially when compared to the distributions of dark gaps from the ionized IGM.  Therefore, we do not expect our main conclusions in this section to be very sensitive to the uncertainties in these parameters.

\section{Discussion}
\label{sec:conc}

Much controversy has surrounded various claims concerning the ionization state of the Universe at $z>6$, as probed by the spectra of high redshift quasars.
  Ironically, the same arguments used to claim that the spectra of these quasars are consistent with reionization completing by $z>>6$ can also be used to claim the spectra are consistent with reionization {\it not} completing by $z<6$.

In this work, we re-examine the often-quoted evidence of $\avenf\lsim10^{-4}$--$10^{-3}$ from the detection of flux in the spectra of $z\lsim6$ QSOs.  Because reionization is very patchy, and the remaining neutral patches become very rare as reionization progresses, it is difficult to claim reionization was complete by this redshift \citep{Lidz07}. 

To quantify these difficulties, we take advantage of the large dynamic range provided by the seminumerical modeling tool, DexM.  Our simulation boxes are 2 Gpc on a side, resolve the likely host halos of QSOs, and generate ionization fields accounting for ionizing photons from halos with virial temperatures of $\sim 10^4$ K.  We extract $\sim 10^6$ ($10^7$) sightlines through the ionization field at $z=$ 6 (5) originating from identified QSO host halos.  However, we do not make detailed simulated spectra for this study.

We confirm that reionization is very inhomogeneous.  In fact, most sightlines through a $\avenf\lsim0.01$ universe do not intersect a single neutral region in a 500 Mpc stretch (corresponding to the redshift interval from $z=6$ to $z=5$. Furthermore, QSO host halos reside in very biased locations of the ionization field, where neutral patches are even rarer. Even without explicitly including the quasar's own ionizing radiation, sightlines originating from QSO halos are over twice as likely than those from random locations to entirely go through the ionized IGM within the first 60--80 Mpc, for $\avenf\lsim0.1$.  As measured by the number density of HI patches, the bias is less extreme, and the asymptotic values are approached at distances of $\sim$ 30--60 Mpc away from the quasar.

Even before overlap, the absorption in the \lya\ forest is dominated by the residual neutral hydrogen inside the ionized IGM at these redshifts.  Detecting the additional dark spectral regions caused by HI patches pre-overlap is very challenging, and cannot be done from the existing \lya\ forest data at $z\sim6$.
At $z\sim5$, where there are more available sightlines and the forest is less dark, constraining $\avenf\lsim0.1$ might be possible given a large dynamic range from very deep spectra and/or the \lyb\ forest.

To further quantify the imprint of an incomplete reionization, we also create dark gap distributions.  In order to be systematic, we separately examine the dark gap PDFs arising from the residual hydrogen in the ionized IGM, and those from just the HI patches pre-overlap.  We find that these two PDFs are likely to have different slopes, which can result in a ``knee'' feature in the combined PDF.  At $z=5$, large dark gaps preferentially come from the HI patches pre-overlap; however, at $z=6$, large dark gaps could instead preferentially come from the ionized IGM.  Unfortunately, averaging the PDFs over redshift bins might smear out these signals, since the gap distribution from the ionized IGM has a strong redshift dependence.

A larger dynamical range would aid in the discriminating power of the current sample of QSO spectra, in both the mean absorption statistic and the dark gap distributions.  An increased dynamic range might be available with new instruments such as the {\it JWST} and the TMT, or through long-integration with existing instruments such as Keck HIRES \citep{BRS07} and ESI, or with higher-order Lyman lines, such as \lyb. In the case of the later, one must be careful in modeling the fluctuating absorption from the superimposed lower redshift \lya\ forest.

%An extended dynamical range can also be achieved with OI, which has a similar ionization potential as HI but with a much weaker optical depth \citep{Oh02_metals}.  Depending on the distribution of metals in the IGM, broad, partially neutral regions pre-overlap could be detectable in deep observations of the OI forest (Cen \& Mesinger, in preparation).

%The Universe could not be mostly neutral at $z\sim5$, since the \lya\ and \lyb\ forest would be darker than observed, and other probes of the neutral fraction sensitive at the level of tens of percent (e.g. Lyman alpha emitters) would be in conflict with observations.  Although an integrated measurement of reionization history, the optical depth to electron scattering determined by {\it WMAP} rules out rapid reionization at $z\lsim6$ by over 2 $\sigma$.  Therefore if indeed the volume weighted neutral fraction of the IGM was a few percent or ten percent at $z\sim5$, reionization must have been a very extended process.  If indeed reionization did finish late and was extended, it would seem to ease the interpretation of a seeming dearth of ionizing photons at $z\sim$ 5--7 as probed by the \lya\ forrest \citep{BH07} and high-redshift Lyman break galaxy surveys (e.g. \citealt{Bouwens08, McLure09, Bouwens09, Ouchi09}).  However, the neutral fractions discussed here are relatively small, and interpreting these observations is very complicated involving many uncertainties.  Therefore, when interpreting these claims, one should not lean too strongly on the crutch of a possible incomplete ionization at $z\sim$5--6.

We propose using the fraction of pixels which are dark as a simple, nearly model-independent upper limit on $\avenf$.  This simple statistic might be sufficient to constrain $\avenf \lsim 0.1$ at $z\sim5$ using deep spectra of the current sample of $z>5$ quasars.

In conclusion, there is currently no direct evidence that reionization was complete by $z\sim$ 5 -- 6.  We stress that we are not promoting such a late reionization scenario.  This paper is merely intended as a caution against over-interpreting high-redshift observations.

\vskip+0.5in

We would like to thank Michael Strauss and David Spiegel for interesting conversations.  We are grateful to Xiaohui Fan and Michael Strauss for providing a list of $z>5$ QSOs.  We also thank Steven Furlanetto, Adam Lidz, Zoltan Haiman, and Renyue Cen for comments on a draft version of this paper.  Support for this work was provided by NASA through Hubble Fellowship grant HST-HF-51245.01-A, awarded by the Space Telescope Science Institute, which is operated by the Association of Universities for Research in Astronomy, Inc., for NASA, under contract NAS 5-26555.

\bibliographystyle{mn2e}
\bibliography{ms}

\end{document}